\documentclass[printer]{aa}  
\usepackage{natbib}                                   
\bibpunct{(}{)}{;}{a}{}{,} 
\usepackage{graphicx}
\usepackage{natbib}

\usepackage{txfonts}
\begin{document}
\authorrunning{Petrovich \and Reisenegger}
\titlerunning{Rotochemical heating in millisecond pulsars with Cooper pairing}
   \title{Rotochemical heating in millisecond pulsars: modified Urca reactions with uniform Cooper pairing gaps.}

   \author{Cristobal Petrovich
                  \and
          Andreas Reisenegger }

   \institute{Departamento de Astronom\'\i a y Astrof\'\i sica, 
		Pontificia Universidad Cat\'olica de Chile,
		Casilla 306, Santiago 22, Chile.
              \email{cpetrovi@astro.puc.cl}}


 
  \abstract
   {When a neutron star's rotation slows down, its internal density increases,
causing deviations from beta equilibrium that induce reactions, heating
the stellar interior. This mechanism, named \textit{rotochemical heating},
has previously been studied for non-superfluid neutron stars.
However, the likely presence of superfluid nucleons will
affect the thermal evolution of the star by suppressing the specific heat
and the usual neutrino-emitting reactions, while at the same time
opening new Cooper pairing reactions.
}
   {We describe the thermal effects of Cooper pairing 
with spatially uniform and isotropic energy gaps of neutrons $\Delta_n$ 
and protons $\Delta_p$, on the rotochemical heating in millisecond pulsars (MSPs) when only 
modified Urca reactions are allowed. In this way, we are able to determine the amplitude
of the superfluid energy gaps for the neutron and protons needed to 
produce different thermal evolution of MSPs.
}
   {We integrate numerically, and analytically in some approximate cases,
 the neutrino reactions for the modified Urca processes with superfluid nucleons 
to include them in the numerical simulation of rotochemical heating.}
 {We find that the chemical imbalances in the star grow up to the 
threshold value $\Delta_{thr}=\mbox{min}\left( \Delta_n+3\Delta_p,3\Delta_n+\Delta_p \right)$,
 which is higher than the quasi-steady
state achieved in the absence of superfluidity. Therefore, the superfluid MSPs
will take longer to reach the quasi-steady state than their nonsuperfluid
counterparts, and they will have a higher a luminosity in this state, given
by $L_\gamma^{\infty,qs} = \left(1-4\right) \times 10^{32}\left(\Delta_{thr}/\mbox{MeV}\right)
\left(\dot{P}_{-20}/P_{\mbox{ms}}^3\right)\mbox{ erg}\mbox{ s}^{-1}$, where
$\dot{P}_{-20}$ is the period derivative in units
of $10^{-20}$ and $P_{\mbox{ms}}$ is the period in milliseconds.
 We are able to explain the UV emission of the PSR J0437-4715 for
$0.05 [ \mbox{MeV}]\lesssim\Delta_{thr}\lesssim0.45 [ \mbox{MeV}]$.
These results are valid if the energy gaps are uniform and isotropic.
}
   {}

\keywords{stars: neutron --- dense matter --- relativity --- stars: rotation
--- pulsars: general --- pulsars: individual (PSR J0437-4715)}

   \maketitle
%

\section{Introduction}
\label{sec:intro}

The observation of thermal emission from the surface 
of a neutron star (NS) has the potential to provide constraints 
on its inner structure. In the existing literature,
several detailed cooling calculations have been compared 
to the few estimates available for the surface temperatures 
of neutron stars (see \citealt{yak04} for a 
review and references). These calculations are based
on passive cooling, at first neutrino-dominated, and later
driven by photon emission at the age of $\sim10^5$yr.

Several mechanisms can keep NSs hot beyond the standard cooling
timescale $\sim10^7$yr. One of them is \textit{rotochemical heating}, which
has previously been studied for neutron stars of non-superfluid matter.
It was first proposed in \citet{reis95} and then developed in \citet{FR05} by considering 
the internal structure via realistic EOSs, in the framework of general relativity.
It works as follows. The reduction in the centrifugal force makes the NS contract. 
This perturbs each fluid element, raising the local pressure and causing deviations 
from beta equilibrium. The system eventually reaches a new quasi-equilibrium configuration where 
the rate at which spin-down modifies the equilibrium concentrations is the same as that at which 
neutrino reactions restore the equilibrium. This implies that rotational energy is converted
into thermal energy and an enhanced neutrino emission is produced by a departure from
the beta equilibrium. Thus, this mechanism can keep old millisecond pulsars (MSPs) warm,
at temperatures $\sim 10^5$ K, which implies that the surface temperature of the
MSP J0437-4715 should be $20\%$ below its measurement (\citealt{FR05}).

On the other hand, cooling curves usually consider the effects of nucleon 
superfluidity  on the thermal evolution of NSs.
Superfluidity is produced by Cooper pairing of baryons 
due to the attractive component of their 
interaction, and it is present only when the temperature $T$
of the matter falls below a critical temperature $T_c$.
The physics of these interactions is rather uncertain 
and very model-dependent, and so is the critical temperature obtained
from theory  (see \citealt{lombardo}).
An important microscopic effect is that the onset of superfluidity
leads to the appearance of a gap $\Delta$ in the spectrum of 
excitations around the Fermi surface. 
This considerably reduces the neutrino reactions and the specific heat 
involving superfluid species \citep{yak01},  
and additionally opens new neutrino emission processes, 
namely pair breaking and formation reactions \citep{flowers}. 
Taking these effects into account, NS cooling has been used 
to constrain the amplitude of the energy gaps by comparing 
predictions with the surface temperatures measured from young neutron stars
(see \citealt{yak04} for a complete discussion). 
In particular, \citet{page} considers the minimal 
cooling paradigm, where no direct Urca processes are allowed and 
the cooling is enhanced by Cooper-pair emission processes.
They find this mechanism to be consistent with observations as long as 
the critical temperature $^3P_2$ of the neutrons
covers a range of values between
$T_c^{min}\lesssim0.2\times10^9$ K and $T_c^{max}\gtrsim0.5\times10^9$ K
in the core of the star.

In this paper, we include the effects of superfluidity in rotochemical 
heating, building on the framework of \citet{FR05} and \citet{reis06}. 
The order of magnitude of these effects was estimated in \citet{reis97}, by
considering the reactions to be totally forbidden, until the chemical imbalances 
$\eta_{npl}= \mu_n- \mu_p- \mu_l$ ( $l=e,\mu$) exceed a 
combination of the energy gaps of either $\Delta_n+\Delta_p$ when direct Urca is allowed or 
$\Delta_n+\Delta_p+2\Delta_<$ when only modified Urca operates, where 
$\Delta_<=\min \left(\Delta_n,\Delta_p \right)$.
This effect lengthens the equilibrium timescale and raises 
the surface temperature.

We consider how exactly the neutrino reactions are suppressed by
superfluidity in the core, by avoiding the step-like approximation of \citet{reis97}.
We calculate instead the reduction in the net modified Urca 
reaction rate and the emissivity
in the regime where the energy gaps, the chemical imbalances, and the temperature are all
relevant quantities. 
In this scenario, \citet{villain} numerically computed  the phase-space
integrals of the net reaction rate for direct Urca and modified Urca
processes, finding that Cooper pairing does not strongly inhibit these
reactions when the energy gaps are not too large compared to both the temperature
and the chemical imbalances. 

The structure of this paper is as follows. In Sect. \ref{sec:theory} we present the basic
equations of rotochemical heating and describe the superfluid input to
compute these numerically. We explain our approach to calculating the reduction factors 
and how they behave in the regime of interest. 
In Sect. \ref{sec:results} we describe our 
results and compare our prediction with the likely thermal emission of PSR J0437-4715
(\citealt{kargaltsev04}) to constrain the values of the energy gaps.
We summarize our main conclusions in Sect. \ref{sec:conclusion}.
 Finally, in Appendices \ref{sec:A} and \ref{sec:B} we explain in detail our numerical
and analytical approaches to computing the net reaction rate and 
emissivity.


		\section{Theoretical framework }\label{sec:theory}

\subsection{Rotochemical heating: basic equations} \label{sec:theory_basic}

The framework of rotochemical heating  used in this work is described 
in detail in \citet{FR05}.
Here we just point out the fundamental equations and the modifications required 
to introduce the effects of superfluidity.
 
We consider the simplest model of a neutron star core, composed
of neutrons, protons, electrons, and muons ($npe\mu$ matter), ignoring
the potential presence of exotic particles.

The internal temperature, redshifted to a distant observer, 
$T_\infty$, is taken to be uniform inside the star because
we are modeling the thermal evolution of a MSP over timescales
much longer than the diffusion time \citep{reis95}. Thus, the evolution of 
the internal temperature for an isothermal interior is given by the thermal 
balance equation \citep{thorne77} 
\begin{eqnarray}
\label{eq:evolucion_Ti}
\dot{T}_\infty & =& \frac{1}{C} \left( L_H^{\infty}-L_\nu^\infty -L_\nu^{BF,\infty}- L_\gamma^\infty\right),
\end{eqnarray}
where $C$ is the total heat capacity of the star,  $L_H^{\infty}$ the
total power released by the heating mechanism, $L_\nu^\infty$ the total 
power emitted as neutrinos due to Urca reactions, $L_\nu^{BF,\infty}$ the total 
power emitted as neutrinos due to Cooper pair-breaking and pair-formation (PB-PF) processes,
and $L_\gamma^\infty$ the power released as thermal photons. 
We briefly discuss the quantities $C$ and $L_\nu^{BF,\infty}$ in 
Sect. \ref{sec:theory_heat} and Sect. \ref{sec:theory_CPemiss}, respectively.  

The amount of energy released by each Urca-type reaction is 
$\eta_{npl}=\mu_n-\mu_p- \mu_l$ ( $l=e,\mu$).
Thus, we write the total energy dissipation rate as
\begin{eqnarray}
\label{eq:heating_term}
L_H^{\infty}= \eta_{npe}^{\infty}\Delta\tilde{\Gamma}_{npe}+\eta_{np\mu}^{\infty}\Delta\tilde{\Gamma}_{np\mu},
\end{eqnarray}
where $\Delta\tilde{\Gamma}_{npl}=\tilde{\Gamma}_{n\rightarrow pl}-\tilde{\Gamma}_{pl\rightarrow n}$ 
is the net reaction rate of the Urca reaction integrated over the core (indicated 
by the tilde) involving the lepton $l$.

The photon luminosity is calculated 
by assuming black-body radiation $L_\gamma^\infty=4\pi\sigma (R^\infty)^2 T_{s,\infty}^4$,
where  $R^\infty$ and $T_{s,\infty}$ are the radius and the surface temperature 
of the star measured from an observer at infinity, respectively. 
To relate the internal and the surface temperatures,
the fully accreted envelope model of \citet{potetal97} is used.

The evolution of the redshifted chemical imbalances, which are also uniform throughout the core, 
are given by
\begin{eqnarray}
\label{eq:evolucion_eta1}
\dot{\eta}^\infty_{npe} &=&  -Z_{npe}\Delta\tilde{\Gamma}_{npe}
- Z_{np}\Delta\tilde{\Gamma}_{np\mu} + 2W_{npe}\Omega \dot{\Omega}\\
\label{eq:evolucion_eta2}
\dot{\eta}^\infty_{np\mu}  &=&  -Z_{np}\Delta\tilde{\Gamma}_{npe}
- Z_{np\mu}\Delta\tilde{\Gamma}_{np\mu} + 2W_{np\mu}\Omega \dot{\Omega},
\end{eqnarray}
where the terms $Z_{np}$, $Z_{npe}$, $Z_{np\mu}$, $W_{npe}$, and $W_{np\mu}$
are constants that depend on the stellar structure and are unchanged 
with respect to their latest definition in \citet{reis06}, and $\Omega \dot{\Omega}$
is the product of the angular velocity and its time
derivative (proportional to the spin-down power). 

The key new contribution in this work is the recalculation 
of $L_\nu^\infty$ and $\Delta\tilde{\Gamma}_{npe}$, 
which differ substantially from those of \citet{FR05}, 
beacuse superfluidity strongly inhibits these reactions.

\subsection{Cooper pairing}
\label{sec:theory_cooper}

In the core, neutrons are believed to form Cooper pairs due to their interaction in the 
triplet $^3P_2$ state, while protons form singlet $^1S_0$ pairs. 
In addition, neutrons in the outermost core and inner crust  are believed to form 
singlet-state $^1S_0$ pairs (type A) \citep{yak01}. 
The $^3P_2$ (type B and C) state description is rather uncertain in the sense 
that the energetically most probable state of $nn$-pairs ($|m_J|=0,1,2$) is not known, 
being extremely sensitive to the still unknown $nn$-interaction
(see, e.g.,  \citealt{amundsen}).
Taking this classification into account, \citet{villain} solve numerically the 
suppression caused by each type of superfluidity of the net reaction rate for 
direct Urca and modified Urca reactions out of beta 
equilibrium, finding that 
the suppression caused by type A superfluidity is in between the suppression by
anisotropic types $|m_J|=0$ (type B) and $|m_J|=2$ (type C)
superfluidity, respectively.
For simplicity, we consider the energy gaps 
for the neutrons $\Delta_n$ and the protons $\Delta_p$ at zero temperature,
 redshifted to a distant observer, as parameters that are 
isotropic ($^1S_0$ pairs) and uniform throughout the core of the NS.

The phase transition for a nucleon species into a superfluid state takes place when
its  temperature falls below a critical value $T_c$. 
This temperature is related to the energy gap at zero temperature $\Delta(T=0)$; 
for the isotropic pairing channel $^1S_0$, 
$\Delta(T=0)=1.764 k T_c$. In addition, when the transition occurs
the amplitude of the energy gap depends on the temperature by means of the 
BCS equation \citep{yak01}, which can be fitted by the practical formula of
\citet{levyak94} for the isotropic gap given by
\begin{eqnarray}
\label{eq:delta}
 \delta \equiv \frac{\Delta(T)}{kT}=\sqrt{1-T/T_c}\left(1.456-\frac{0.157}{\sqrt{T/T_c}}+\frac{1.764}{T/T_c}\right),
\end{eqnarray}
where $\delta$ is the variable used in the phase-space integrals in
Sect. \ref{sec:theory_emiss} that depends on the input parameters $\Delta_n$ and $\Delta_p$  used throughout this paper.
It is straightforward to check that the limiting cases are reproduced by the Eq.
(\ref{eq:delta}), i.e., $\delta=0$ when $T=T_c$ and $\delta=\Delta(T=0)/kT$ when $T\ll T_c$. 
In addition, \citet{levyak94} claim that intermediate values of $T/T_c$ are also
reproduced by this formula with a maximum error smaller than $5\%$, which is accurate
enough for the purpose of this work.

Having defined the energy gap of the nucleon $\Delta_i$ with $i=n,p$, it is possible 
to express the momentum dependence of the nucleon energy $\epsilon_i(p_i)$ near the Fermi
level, i.e. $|p_i-p_{F_i}|\ll p_{F_i}$, as in \citet{yak01} as
\begin{eqnarray}
\label{eq:dispertion}
 \epsilon_i(p_i)=\mu_i- \sqrt{v_{F_i}^2(p_i-p_{F_i})^2+\Delta_i^2}\quad\mbox{if}\quad p_i<p_{F_i} ,\nonumber\\ 
 \epsilon_i(p_i)=\mu_i+ \sqrt{v_{F_i}^2(p_i-p_{F_i})^2+\Delta_i^2}\quad\mbox{if}\quad p_i>p_{F_i},
\end{eqnarray}
where $p_{F_i}$, $v_{F_i}$, and  $\mu_i$ are the Fermi momentum, the Fermi velocity,
and the chemical potential of species $i=n,p$, respectively.

\subsection{Neutrino emissivity}
\label{sec:theory_emiss}

The most rapid reactions in NS cores are the direct 
Urca processes. However, as already mentioned, we assume in this work that 
these reactions are forbidden because direct Urca with superfluidity might drastically 
change the behavior of rotochemical heating, leading to new conclusions that will 
be discussed in a forthcoming work.
Additionally, these reactions can be kinematically forbidden for a wide range of 
EOSs and central densities.
If this were the case, the so-called modified Urca reactions (or Murca reactions) 
would prevail \citep{yak01} because they involve
an additional spectator nucleon $N$ that allows energy and momentum conservation.
In general, these reactions are
\begin{eqnarray}
n+N_i\rightarrow p+N_f+e^-+\bar{\nu}_e \\
p+N_i+e^-\rightarrow n+N_f+\nu_e, 
\end{eqnarray}
where the subindices $i$ and $f$ represent the
for its initial and final states of the spectator nucleon. If this nucleon is a proton, 
these reactions are called the \textit{proton branch} of Murca, and 
if it is a neutron, they are called the \textit{neutron branch} of Murca.

We write the neutrino emissivity and the net 
reaction rate due to Murca reactions involving the lepton $l$ and 
 integrated over the core, respectively, as
\begin{eqnarray}
\label{eq:L_noneq}
L_{\nu,l}^\infty & = & \tilde{L}_{nl}I_{M,\epsilon}^n T_\infty^8 +\tilde{L}_{pl}I_{M,\epsilon}^p T_\infty^8,\\
\label{eq:gamma_noneq}
\Delta\tilde{\Gamma}_{npl} & = & \frac{\tilde{L}_{nl}}{k}I_{M,\Gamma}^n T_\infty^7 +\frac{\tilde{L}_{pl}}{k}I_{M,\Gamma}^p T_\infty^7,
\end{eqnarray}
where constants $\tilde{L}_{nl}$ and $\tilde{L}_{pl}$ are defined in terms 
of the neutrino luminosities for a nonsuperfluid NS in beta equilibrium, as
\begin{eqnarray}
\label{eq:int_Qeq}
\tilde{L}_\alpha & \equiv & \frac{L_\alpha^{eq}}{T_\infty^8} = \int_{\mbox{core}} 4\pi r^2 e^{\Lambda} S_\alpha(n) e^{-6\Phi} dr,
\end{eqnarray}
where $\alpha$ indicates both the branch of the Murca process and the lepton involved
in the reaction \citep{FR05}. The term $S_\alpha$ is a slowly varying function of the 
baryon number  density $n$ (e.g., \citealt{yak01}), and $\Lambda$ and $\phi$ are the usual 
Schwarzschild metric terms. 
The quantities $I_{M,\epsilon}^N$ and $I_{M,\Gamma}^N$
are dimensionless phase-space integrals 
that contain the dependence of the emissivity and the net reaction rate, 
respectively, on the chemical imbalances $\eta_{npl}^\infty$ and on the 
energy gaps $\Delta_n$ and $\Delta_p$.

To introduce these integrals, it is useful to define the
usual dimensionless variables normalized by the thermal energy $kT$, of
\begin{equation}
\label{eq:adim_var_1}
 x_j\equiv \frac{\epsilon_j-\mu_j}{kT} , \mbox{  }\mbox{  } x_\nu\equiv \frac{\epsilon_\nu}{kT},
\mbox{  }\mbox{ and } \mbox{  } \xi_l\equiv \frac{\eta_{npl}}{kT}  ,
\end{equation}
which represent the energy of the nonsuperfluid degenerated particle $j$, 
the neutrino, and the chemical imbalance involving
the lepton $l$, respectively, while for the superfluid nucleon $i$ we write
\begin{equation}
\label{eq:adim_var_2}
 x_i\equiv\frac{v_{F_i}(p_i-p_{F_i})}{kT}  \mbox{ and } 
z_i \equiv \mbox{sgn}(x_i)\sqrt{x_i^2+\delta_i^2}.
\end{equation}
In terms of these variables,
\begin{eqnarray}
\label{eq:integral_murca}
I_{M,\epsilon}^N&=&\frac{60480}{11513\pi^8}\cdot \int_0^{\infty}dx_{\nu} x_{\nu}^3 \int_{-\infty}^{\infty}\int_{-\infty}^{\infty}\int_{-\infty}^{\infty}\int_{-\infty}^{\infty}\int_{-\infty}^{\infty}
dx_{n}dx_{N_i}dx_{N_f}  \nonumber\\
&&dx_{p}dx_{e}f(z_n)f(z_{N_i})f(z_{N_f}) f(z_p)f(x_e) \nonumber \\
&&\times \left[\delta(x_{\nu}+\xi_l-z_n-z_{N_i}-z_{N_f}-z_p-x_e) \right.  \nonumber\\
&&+\left.\delta(x_{\nu}-\xi_l-z_n-z_{N_i}-z_{N_f}-z_p-x_e) \right]
\end{eqnarray}
and
\begin{eqnarray}
\label{eq:integral_murca_gamma}
I_{M,\Gamma}^N&=&
\frac{60480}{11513\pi^8}\cdot\int_0^{\infty}dx_{\nu} x_{\nu}^2 \int_{-\infty}^{\infty}\int_{-\infty}^{\infty}\int_{-\infty}^{\infty}\int_{-\infty}^{\infty}\int_{-\infty}^{\infty}
dx_{n}dx_{N_i}dx_{N_f}  \nonumber\\
&&dx_{p}dx_{e}f(z_n)f(z_{N_i})f(z_{N_f}) f(z_p)f(x_e) \nonumber \\
&&\times \left[\delta(x_{\nu}+\xi_l-z_n-z_{N_i}-z_{N_f}-z_p-x_e) \right.  \nonumber\\
&&-\left.\delta(x_{\nu}-\xi_l-z_n-z_{N_i}-z_{N_f}-z_p-x_e) \right],
\end{eqnarray}
where  $f(\cdot)$ is the Fermi function $f(x)=1/(1+e^x)$ and the numerical
factor in front of the integral is to normalize it to 1  when the
energy gaps and the chemical imbalances are zero.

In the nonsuperfluid case (i.e.,  $\delta_n=\delta_p=0$), 
these integrals reduce to 
the polynomials calculated by \citet{reis95}
\begin{eqnarray}
\label{eq:F_M_def}
&&I_{M,\epsilon}^N (\delta_n=\delta_p=0)=F_M(\xi_l)= \nonumber\\
&&\quad \quad  1 + \frac{22020\xi_l^2}{11513\pi^2} + \frac{5670\xi_l^4}{11513\pi^4} + \frac{420\xi_l^6}{11513\pi^6} + \frac{9\xi_l^8}{11513\pi^8},\\
\label{eq:H_M_def}
&&I_{M,\Gamma}^N (\delta_n=\delta_p=0)=H_M(\xi_l)= \nonumber\\ 
&&\quad \quad\frac{14680\xi_l}{11513\pi^2}+\frac{7560\xi_l^3}{11513\pi^4}+\frac{840\xi_l^5}{11513\pi^6} +\frac{24\xi_l^7}{11513\pi^8},
\end{eqnarray}
which are the same for the neutron branch and the proton branch.

Since the $z$-variables defined in Eq. (\ref{eq:adim_var_2}) depend on the energy gaps,
the equality between $I_{M,\epsilon}^n$ and $I_{M,\epsilon}^p$, or between
 $I_{M,\Gamma}^n$ and $I_{M,\Gamma}^p$, is no longer satisfied
if the gaps are different. 

\subsubsection{Reduction factors}
\label{sec:theory_emiss_red}

The phase-space integrals in Eqs. (\ref{eq:integral_murca}) and 
(\ref{eq:integral_murca_gamma}) do not have an analytical expression when
one or both of the reacting nucleons are superfluid. Thus, their calculation
must be done numerically, as in \citet{villain}. A natural way
to account for the suppression produced by Cooper pairing is to define 
the so-called reduction factors as the ratio of these superfluid integrals
to their non-superfluid limits
\begin{eqnarray}
\label{eq:R_epsilon}
 R_{M,\epsilon}^N(\xi_l,\delta_n,\delta_p)&=& I_{M,\epsilon}^N(\xi_l,\delta_n,\delta_p)/F_M(\xi_l),\\
\label{eq:R_gamma}
 R_{M,\Gamma}^N(\xi_l,\delta_n,\delta_p)&=& I_{M,\Gamma}^N(\xi_l,\delta_n,\delta_p)/H_M(\xi_l).
\end{eqnarray}
In principle, to calculate these reduction factors, a five-dimensional 
integral needs to be computed, because only one dimension can be eliminated by
integrating out the electron variable in Eqs.
(\ref{eq:integral_murca}) and (\ref{eq:integral_murca_gamma}). 
On the other hand, if one of the nucleons is not superfluid, one can eliminate 
more dimensions by integrating out the nonsuperfluid variables. 
For instance, if we consider both the neutron branch and the protons as the only 
superfluid  species, we can integrate out the electron plus the three 
nonsuperfluid neutrons, obtain a two-dimensional integral that has to be 
calculated numerically (see \citealt{villain} for the formulae and details).
However, it may be a problem to calculate these integrals efficiently including
the new superfluid luminosities and net reaction rates in the rotochemical evolution
equations.

The main difficulties in performing the integration
of Eqs. (\ref{eq:integral_murca}) and (\ref{eq:integral_murca_gamma}) are
\begin{itemize}
\item infinite sizes of the integration domains;
\item external free parameters $\eta_l$, $\delta_n$, and $\delta_p$, which can be  very large;
\item many integration dimensions (up to five).
\end{itemize}

\citet{villain} calculate these integrals  numerically
via the so-called Gauss-Legendre quadrature, using
logarithmic variables scaled to external parameters $\eta_l$, $\delta_n$, $\delta_p$. 
In this way they cover a wide range 
of the infinite integration domain, 
for the parameters in Eq. (\ref{eq:R_gamma})
in the range $\delta_N \in [0,10^3]$ and 
$\xi_l\in[0,10^4]$. 

We require a fast integration method because we need to evaluate 
eight integrals (emissivity and net
reaction rate for two nucleon branches and two leptons branches) for each time-step
of the thermal evolution. For this reason, we used a slightly different
method than that of \citet{villain}, although we also implemented
their code to calibrate ours.
We chose the Gauss-Laguerre method, which is accurate enough for 
a few evaluations when  the integrands are asymptotically exponentially decaying 
functions, as is the case for the Fermi functions 
(see Appendix \ref{sec:A} for a detailed explanation).

One striking feature of the reduction factors is that, for relatively high 
values of the energy gaps and chemical imbalances compared to the thermal scale,
they tend to be independent of the temperature.
\citet{villain}, considering only one superfluid particle species,
claim by graphical inspection that, when $\eta\gtrsim\Delta$ and $\Delta>10kT$, 
the  reduction factors become functions of $\Delta/\eta$ only.
By using our numerical results, we show in Appendix B that
this is a good approximation if $\Delta\gtrsim30kT$ and $\eta>\Delta$.
In the limit of zero temperature and in the presence of one superfluid 
variable we find the analytical expressions
(see Appendix \ref{sec:B} for more details)
\begin{eqnarray}
\label{eq:R_epsilon_aprox}
R_{M,\epsilon}(\Delta/\eta)&=&\left( 28 (\Delta/\eta)^2+ 105 (\Delta/\eta)^4 + \frac{105}{2}(\Delta/\eta)^6 + \frac{35}{16}(\Delta/\eta)^8\right)
\nonumber \\
&\times& \ln\left(\frac{\Delta/\eta}{1+\sqrt{1-(\Delta/\eta)^2}}\right) + \sqrt{1-(\Delta/\eta)^2}\nonumber\\ 
&\times& \left(1+\frac{551}{10}(\Delta/\eta)^2 +\frac{4327}{40}(\Delta/\eta)^4+ \frac{1873}{80}(\Delta/\eta)^6\right)\\
\label{eq:R_gama_aprox}
R_{M,\Gamma}(\Delta/\eta)&=&\left(21(\Delta/\eta)^2+ \frac{105}{2}(\Delta/\eta)^4 + \frac{105}{8}(\Delta/\eta)^6\right)
\nonumber \\
&\times& \ln\left(\frac{\Delta/\eta}{1+\sqrt{1-(\Delta/\eta)^2}}\right) + \sqrt{1-(\Delta/\eta)^2}\nonumber\\ 
&\times&\left(1+\frac{759}{20}(\Delta/\eta)^2 +\frac{1779}{40}(\Delta/\eta)^4+\frac{16}{5}(\Delta/\eta)^6\right),
\end{eqnarray}
where $\Delta$ is the energy gap of one of the nucleons, i.e., the proton if we 
consider the neutron branch and the neutron for the proton branch.
It is straightforward to verify that these expressions satisfy the limiting cases
$\lim_{\Delta/\eta\rightarrow0}R=1$ and $\lim_{\Delta/\eta\rightarrow1}R=0$
as expected. In addition, in the limit $kT\ll1$ the functions in Eqs. (\ref{eq:F_M_def})
and (\ref{eq:H_M_def}) tend to the highest power of the polynomial, i.e., 
$F(\xi)\cong\frac{9\xi^8}{11513\pi^8}$ and
$H(\xi)\cong\frac{24\xi^8}{11513\pi^8}$. Thus, the integrals can be expressed as
 \begin{eqnarray}
I_{\epsilon}(\delta,\xi)&=&\frac{9\xi^8}{11513\pi^8}R_{\epsilon}(\delta/\xi), \mbox{and} \\
I_{\Gamma}(\delta,\xi)&=&\frac{24\xi^8}{11513\pi^8}R_{\Gamma}(\delta/\xi), 
\end{eqnarray}
where, doing some algebra, one can verify that 
these approximations satisfy the identity of \citet{flores}, i.e., 
 $\frac{dI_{\epsilon}(\delta,\xi)}{d\xi}=3I_{\Gamma}(\delta,\xi)$. 
In Appendix \ref{sec:B}, we compare these formulae with the numerical 
non-zero temperature calculations, finding that they agree to a very good approximation
when $\Delta/T\gtrsim 30$.

\begin{figure}[!h]
\includegraphics[width=9cm,height=5.5cm]{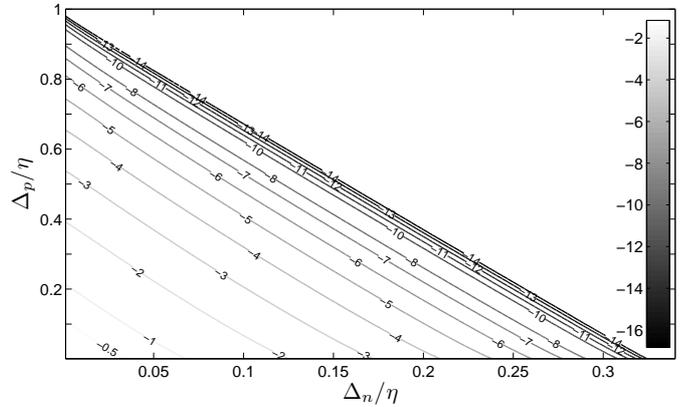} 
\caption{
Contour plot of the reduction factor in the 
net reaction rate in the neutron branch of Murca 
in the limit of zero temperature. The numbers are 
$\log(R_{M,\Gamma}^n)$.}
\label{fig:1}
\end{figure}

When two nucleons are superfluid, we cannot find an analytical 
expression, but we can simplify the problem to a three-dimensional,
bounded integral. This method is obviously much faster than the finite temperature
one with the quadrature integration in Appendix \ref{sec:A}. 
However, it is an approximation to the problem, 
and only works when the temperature is quite low compared to the chemical imbalance 
and energy gaps.
Furthermore, if we analyze the case with two superfluid
nucleons, the limit works when the chemical imbalance overcomes a certain 
energy threshold $\Delta_{thr}$ imposed by the energy gaps $\Delta_n$ and $\Delta_p$. 
If we consider the reaction $n+n_i\rightarrow n_f+p+e^-+\bar{\nu}_e$ 
at zero temperature, the neutron is energetically allowed to decay only when 
$2\mu_n-2\Delta_n>\mu_n+\Delta_n+\mu_p+\Delta_p+\mu_e$, which implies
the simple condition $\eta_{npe}>3\Delta_n+\Delta_p$ (see \citealt{reis97}
for a schematic justification of this).
Finally, the threshold imposed by the gaps $\Delta_{thr}$ for both branches
of Murca reactions are
\begin{eqnarray}
\label{eq:thr_neutron}
 \Delta_{thr}&=& 3\Delta_n+\Delta_p \quad \mbox{for the neutron branch and},\\
\label{eq:thr_proton}
 \Delta_{thr}&=& \Delta_n+3\Delta_p \quad \mbox{for the proton branch}.
\end{eqnarray}

In Fig.~\ref{fig:1}, we plot the reduction factor in the net reaction
rate of the neutron branch of Murca under the approximation of zero temperature. 
As shown in this figure, there are no reactions when $\eta<\Delta_{thr}=3\Delta_n+\Delta_p$.
Remarkably, the contour lines of the reduction factor are almost straight lines 
that coincide with the slope of the contour levels of equation
$3\Delta_n+\Delta_p$. This means that the reduction
factor in this regime is close to a certain function of $\Delta_{thr}$, which
is valid for both branches and also for the reduction factor of the emissivity.

\begin{figure}[!h]
\includegraphics[width=8.8cm,height=5.5cm]{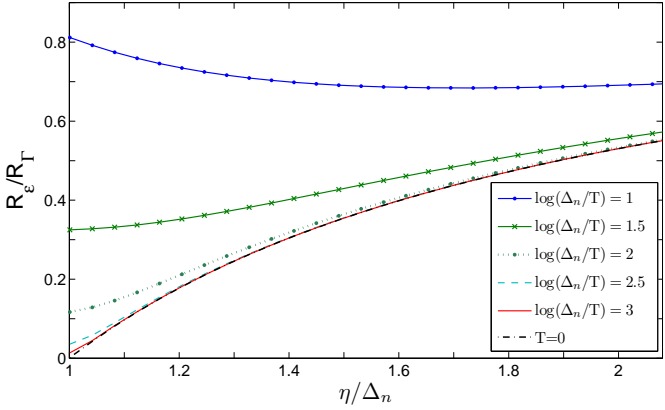} 
\caption{
Ratio of the reduction factor of the emissivity $R_\epsilon$ to
the reduction factor of the net reaction rate $R_\Gamma$ for the
proton branch of Murca reactions with superfluid neutrons
as a function of $\eta/\Delta_n$ for different values of $\Delta_n/T$.
Ratio of the zero temperature approximations given by
Eqs. (\ref{eq:R_epsilon_aprox}) and (\ref{eq:R_gama_aprox})
is given by the black dash-dotted line.
}
\label{fig:2}
\end{figure}

Since we obtain the reduction factors in the 
net reaction rate that are similar to those found by \citet{villain}, 
these can be seen in their work. 
The behavior of the reduction factor in the emissivity is also similar to 
that in the net reaction rate. Thus, to illustrate the difference between these two quantities, 
we plot  their ratio $R_\epsilon/R_\Gamma$ in Fig.~\ref{fig:2} for the range
where the chemical imbalance slightly exceeds the energy gaps for different
values of the temperature compared to the gaps. 
We show in Sect. \ref{sec:results} that this is the regime of interest for rotochemical heating. 
We can verify using the formulae in Eqs. (\ref{eq:R_epsilon_aprox})
and (\ref{eq:R_gama_aprox}) that the ratio $R_\epsilon/R_\Gamma$ approaches a value of one 
when $\eta/\Delta\rightarrow\infty$.
The ratio  also decreases as the temperature diminishes compared to the 
energy gap until the limiting case of zero temperature. This is even more dramatic if
we have two superfluid particle species.
In conclusion, for very low temperatures relative to the other energy scales 
and chemical imbalances slightly
above the energy gap, the suppression of the emissivity is much higher than that
of the net reaction rate. 
Physically what happens is that the neutrino released
in the reaction escapes with a very small amount of energy, which at zero temperature
is proportional to the excess of energy $\sim(\eta-\Delta_{thr})$ that can be 
arbitrarily small as $\eta$ approaches $\Delta_{thr}$.

We can finally incorporate the recalculation of the emissivity and net reaction
rate into the rotochemical equations when one of the nucleon species is superfluid,
by applying the Gauss-Laguerre quadrature method to the entire evolution.
In contrast, when two superfluid nucleons are present, we separate the evolution 
into two regimes and follow their respective approaches. While $\eta<\Delta_{thr}$, we use the
Gauss-Laguerre quadrature method explained in Appendix \ref{sec:A}, which happens for 
the early evolution of rotochemical heating (see Fig.~\ref{fig:3}). 
When $\eta>\Delta_{thr}$, we use the analytical approximation of the integral presented
in Appendix \ref{sec:B}.

\subsection{Specific heat suppression}
\label{sec:theory_heat}

When $T$ reaches $T_c$, there is a discontinuous 
increase in the specific heat that is characteristic of a second order phase
transition. \citet{yak01} claim that this increase is by a factor of $\sim 2.4$ with respect
to the normal-matter specific heat.
Subsequently, when $T\ll T_c$ , an exponential-like suppression 
occurs because of the gap in the energy spectrum. 
In practice, these effects are taken into account by
using control functions that  multiply the unpaired values of the 
specific heat at constant volume $C_V$. 
These have been calculated by several authors
for various temperature regimes, e.g. \citet{pizzochero} for $T\ll T_c$ and \citet{maxwell}
for $0.2T_c<T<T_c$. The former set of authors obtain an exponential decaying control function of the
form $\frac{\Delta}{kT}e^{-\Delta/kT}$ and the latter, which is the result we adopt, obtain
the function
\begin{eqnarray}
\label{eq:c_V_sup}
C_{V,N}^{sup}&=&C_{V,N}\times3.15\frac{T_{c,N}}{T}e^{-1.76T_{c,N}/T}\times \nonumber\\
&& \times \left[2.5-1.66 \frac{T}{T_{c,N}} +3.64
\left(\frac{T}{T_{c,N}}\right)^2  \right],
\end{eqnarray}
where $T_{c,N}$ is the critical temperature of the nucleon $N$ and 
$C_{V,N}$ is the specific heat of the normal matter case, as
defined in \citet{FR05}.
As might be noticed, this expression also repruduces the result of \citet{pizzochero} in the 
low-temperature regime. The value at the lower limit of the 
expression found by \citet{maxwell} is indeed $C_{V,N}^{sup}(T=0.2T_c)/C_{V,N}=0.005$, 
which is sufficiently small to ensure a 
the lepton contribution to the specific heat dominant. 
This formula also represents the discontinuous increase 
at $T=T_c$, which implies that  $C_{V,N}^{sup}(T=T_c)/C_{V,N}=2.42$, in agreement with \citet{yak01}.
Finally, an important remark is that the minimum value that the specific heat
can reach  is the leptonic contribution, which is $C_V=C_{V,e}+C_{V,\mu}$.

\subsection{Cooper pairing emission}
\label{sec:theory_CPemiss}

Another feature of the superfluid state is the appearance of 
new neutrino reaction mechanisms due to  the Cooper pairs. These are the
Cooper pair breaking and pair formation processes proposed by 
\citet{flowers}. These authors
claim that a superfluid neutron star can be considered as a two-component system, 
which consists of paired quasiparticles and 
elementary excitations above the condensate.
Their associated quasi-equilibrium densities are controlled by the processes, which are 
prevalent at temperatures close to $T_c$ and successively suppressed at lower
temperatures because of an exponential decrease in the number of unpaired particles.
Schematically, these neutrino reactions are
\begin{eqnarray}
 {NN}&\rightarrow& N+N+\nu+\bar{\nu} \quad  \mbox{ pair breaking (PB)},\\
  N+N&\rightarrow& {NN}+\nu+\bar{\nu} \quad  \mbox{ pair formation (PF)},
\end{eqnarray}
where ${NN}$ denotes the Cooper pair and $N$ the excitation.\\
These authors find an emissivity 
 $ Q_{PB,PF}\sim 10^{28}(kT/\mbox{[MeV]})^7$ $\left[\mbox{erg}\mbox{ cm}^{-3}\mbox{ s}^{-1}\right]$
 when $T \lesssim T_c$ . 
Thus, these reactions could certainly affect the thermal evolution of the early stage, since
the order of magnitude of the emissivity for the Murca processes is 
$Q_{MU}\sim 10^{29-30} (kT/\mbox{[MeV]})^8 \left[\mbox{erg}\mbox{ cm}^{-3}\mbox{ s}^{-1}\right]$.
However, the relevant question here is whether or not it affects the 
late stage, i.e., when the photon luminosity is the dominant cooling
mechanism.
In this sense, we estimate the PB-PF emissivity for a
a typical scenario of rotochemical heating, where the temperature
that it reaches in the late quasi-steady state  is $kT\sim10^{-3}$ MeV
(see \citealt{FR05}). 
In this limit, in which $T\ll T_c$ \citep{flowers},
\begin{eqnarray}
 Q_{PB,PF}\sim 10^{28}\left(\frac{\Delta}{\mbox{MeV}}\right)^7\sqrt{\Delta/kT}e^{-2\Delta/kT} \mbox{ }
\mbox{ erg}\mbox{ cm}^{-3}\mbox{ s}^{-1},
\end{eqnarray}
which clearly becomes negligible when the values of the energy gaps
are relatively large compared to the thermal scale, say $\Delta/kT\sim10^{2}-10^3$,
because the exponential behavior rapidly suppresses the effects of these reactions.
For $\Delta>0.01$ MeV,  we check that the PB-PF processes become irrelevant compared
to the photon luminosity, and therefore 
do not contribute to the total luminosity at this stage.

		\section{Results and discussion}\label{sec:results}


\subsection{Evolution}\label{sec:results_evol}

We show the results of numerically solving the rotochemical evolution
Eqs. (\ref{eq:evolucion_Ti}), (\ref{eq:evolucion_eta1}), and (\ref{eq:evolucion_eta2}),
considering the inputs in Sect. \ref{sec:theory}. 

Hereafter, we model the neutron star structure by the 
 A18 + $\delta \upsilon$ + UIX* equation of state (EOS) \citep{apr98}.
The most relevant feature of this EOS in this work is that it allows 
direct Urca reactions for electrons above a density $\rho_D=1.59\times10^{15}$ g cm$^{-3}$,
which corresponds to the central density of a 2 $M_\odot$ star. On the other hand,
the threshold for direct Urca reactions of muons lies at a higher density 
in a non-causal regime. We consider stars below this mass limit, in which
no direct Urca processes occur.

\begin{figure}[!h]
\includegraphics[width=9cm,height=5.5cm]{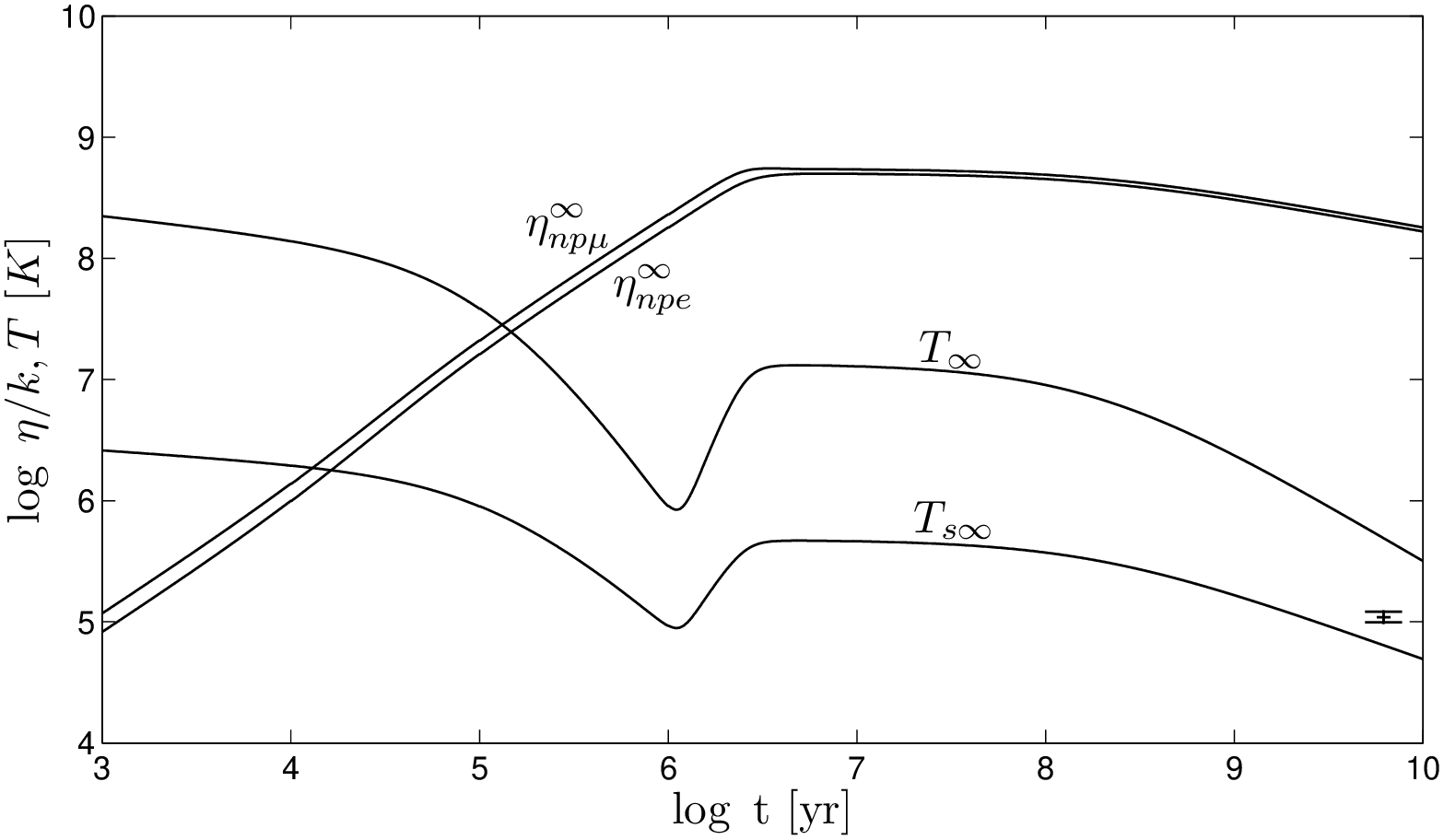}
\includegraphics[width=9cm,height=5.5cm]{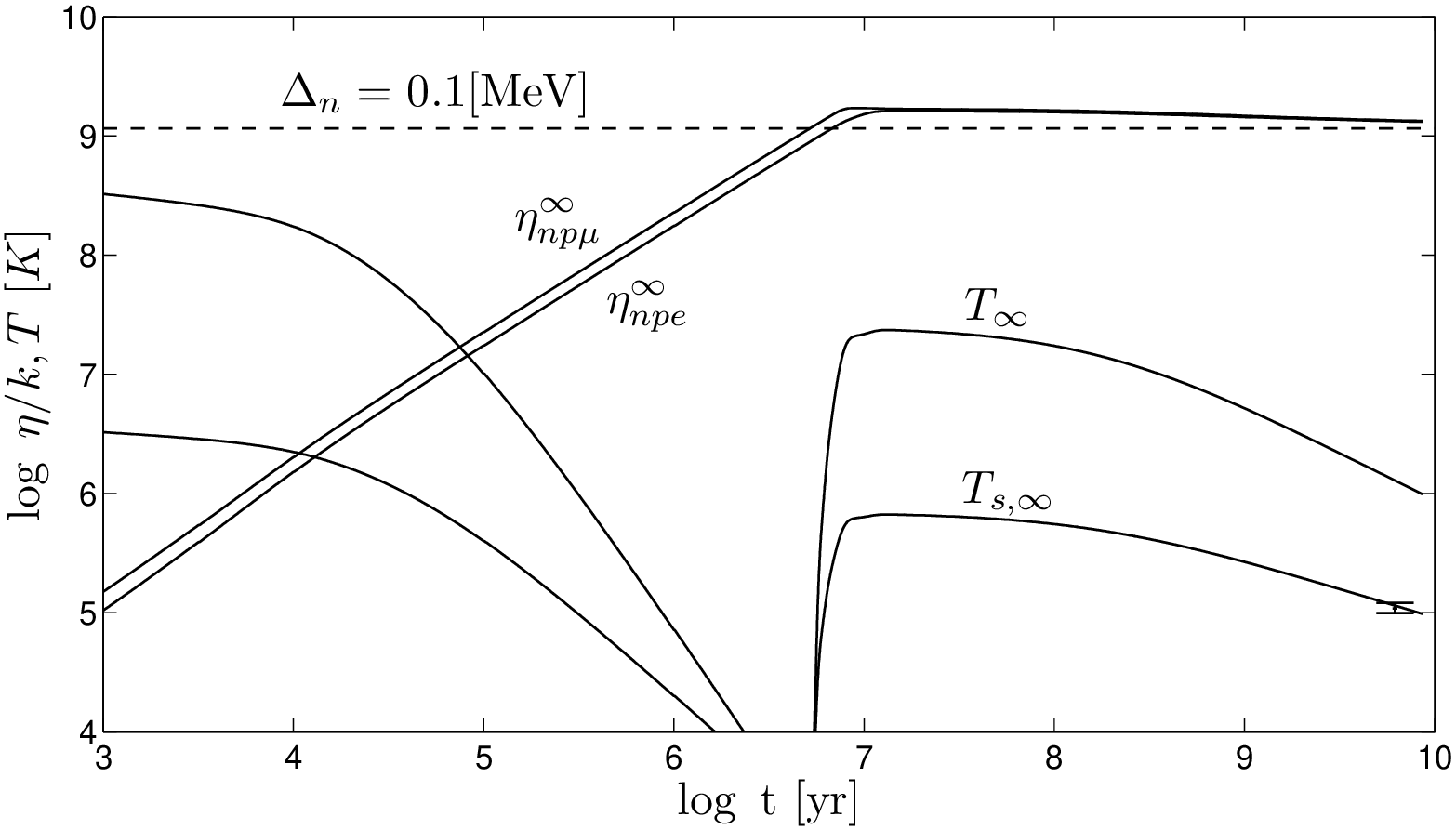} 
\caption{
Evolution of the internal temperature $T_\infty$, the surface temperature $T_{s,\infty}$ and the
chemical imbalances $\eta_{npe}^\infty$, $\eta_{npu}^\infty$ for a 
star with the mass fixed to the PSR J0437-4715, i.e.
 $1.76 M_{\odot}$ \citep{verbiest}, built with the 
A18 + $\delta\upsilon$ + UIX* EOS, with the initial condition $T_{\infty}=10^9$ K 
and null chemical imbalances. The spin-down  is assumed to be due to the 
magnetic dipole radiation, with dipolar field strength
 $B=2.8 \times 10^8$ G and 
initial period of $P_0=1$ ms. The error bar is the $90\%$ confidence level for 
the surface temperature measured for the PSR J0437-4715 \citep{kargaltsev04} at the 
current spin-down parameters.
\textit{Upper panel:} nonsuperfluid case (null energy gaps). 
\textit{Lower panel:} superfluidity of neutrons with $\Delta_n=0.1$ MeV (dashed line).}
\label{fig:3}
\end{figure}

The results are shown in Fig.~\ref{fig:3}, where
in the upper panel we plot the solution to this set of equations for the 
nonsuperfluid case, and in the lower we compare these to our results considering
superfluidity of neutrons for a constant energy gap of $\Delta_n=0.1$ MeV
by integrating numerically the emissivity and net reaction rate without
any approximation.

Except for the superfluidity energy gap of the neutrons in the 
lower panel, both panels in Fig.~\ref{fig:3} have the same input parameters. 
This figure indicates that during the neutrino cooling stage, 
the differences between  the nonsuperfluid and superfluid 
cases are not noticeable. 
After $10^5$ yr, the temperature drops more rapidly with time in the 
superfluid case because the specific heat of the
neutrons is suppressed. The photon cooling, which
starts to dominate in both cases, is unaltered by the presence
of the Cooper pairs. 

Later on, the nonsuperfluid case  reaches a \textit{quasi-steady} state, 
in which the rate at which the spin-down modifies 
equilibrium concentrations is the same as the rate at which the reactions
drive the system toward the equilibrium concentration, with heating
and cooling also balancing each other (see \citealt{reis95} and \citealt{FR05} for more
details). 
The timescale on which the superfluid case reaches this quasi-steady
state is longer because the Murca reactions are highly suppressed when the chemical
imbalances are smaller than $\Delta_n$. But, immediately after the chemical imbalances
overcome the value of the energy gap of the neutrons, several neutrino reactions
are produced, which drastically reheat the star, causing a
rapid increase in the temperature to finally reach the quasi-steady state. 
The subsequent evolution 
can be approximated by the simultaneous solution of Eqs.
(\ref{eq:evolucion_Ti}), (\ref{eq:evolucion_eta1}), and (\ref{eq:evolucion_eta2})
with their left-hand sides set equal zero, as discussed in Sect. \ref{sec:results_qs}.
Finally, the chemical imbalances at this final evolution stage reach a higher value 
than in the nonsuperfluid case, lengthening its timescale to reach
quasi-steady state. This implies that the
star can store more chemical energy and it is released it later in the 
time evolution compared to the nonsuperfluid case.
Therefore, the heating when Cooper pairing is present is more efficient
to ensure that the MSP at higher temperatures during the quasi-steady state,
 compared to the normal matter case.
Moreover, the choice of superfluid of neutrons with $\Delta_n=$0.1 MeV and nonsuperfluid
protons can explain the 90$\%$ confidence level of the 
surface temperature of the MSP J0437-4715 \citep{kargaltsev04},
 which the nonsuperfluid case cannot.

This aforementioned effect of superfluidity on rotochemical heating was already
predicted by \citet{reis97} via a rough estimation; this author claimed that the neutrino 
reactions opened when they overcome the threshold $\Delta_{thr}$, that for Fig.~\ref{fig:3} is
$\Delta_{thr}=\Delta_n$ (see Sect. \ref{sec:theory_emiss_red}), 
ignoring the temperature dependence of these reactions.
By looking at in Fig.~\ref{fig:3} and solving the evolution equations
for several combinations of $\Delta_n$ and $\Delta_p$ in the range 
of $\Delta_{thr}=0.05-1$ MeV,
we verify that for the Murca processes, the zero temperature approach
of \citet{reis97} is valid, because in the quasi-steady state
$\eta_{npe},\eta_{np\mu}\gtrsim\Delta_{thr}$.
To a good approximation, this means that the net reactions rates and the emissivity
indeed do not depend on the internal temperature, as we showed in 
Sect. \ref{sec:theory_emiss_red}.

\subsection{Quasi-steady state}\label{sec:results_qs}
 
\begin{figure}[!h]
   \centering
\includegraphics[width=9.3cm,height=5.8cm]{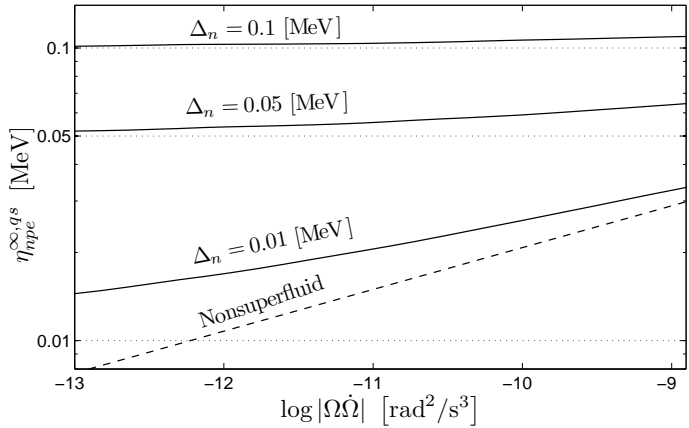} 
\caption{ Chemical imbalance of the electrons in the quasi-steady state
with only neutrons as superfluid  particles as a function of  $\Omega \dot{\Omega}$
for three values of the energy gap $\Delta_n$ 
 (solid lines) and the nonsuperfluid case (dashed line) 
for a $1.76 M_{\odot}$ star calculated with the 
A18 + $\delta\upsilon$ + UIX* EOS. The horizontal lines
indicate the values of the energy gaps: $0.01,0.05$, and $0.1$ MeV (dotted lines). 
}
\label{fig:4}
\end{figure}

Sufficiently old stars will have reached the quasi-steady state, 
where  $\dot{T}_\infty=0$, $\dot{\eta}_{npe}^{\infty}=0$, 
and $\dot{\eta}_{np\mu}^{\infty}=0$.
In this state, the Eqs. (\ref{eq:evolucion_Ti}), (\ref{eq:evolucion_eta1}),
and (\ref{eq:evolucion_eta2}) can be written as
\begin{eqnarray}
\label{eq:equilibrio_Ti}
L_\gamma^{\infty,qs}& =&  \eta_{npe}^{\infty,qs}\Delta\tilde{\Gamma}_{npe}+\eta_{np\mu}^{\infty,qs}\Delta\tilde{\Gamma}_{np\mu}- L_\nu^\infty, \\
\label{eq:equilibrio_eta1}
2  W_{npe}\Omega \dot{\Omega} &=&  Z_{npe}\Delta\tilde{\Gamma}_{npe}+
 Z_{np}\Delta\tilde{\Gamma}_{np\mu}, \\
\label{eq:equilibrio_eta2}
2  W_{np\mu}\Omega \dot{\Omega} &=&  Z_{np}\Delta\tilde{\Gamma}_{npe}+
 Z_{np\mu}\Delta\tilde{\Gamma}_{np\mu},
\end{eqnarray}
where the superscript $qs$  stands for quasi-steady
and we have neglected the neutrino contribution
of the Cooper pairing emission as explained in Sect. \ref{sec:theory_CPemiss}.
This system of equations totally specifies the temperature and
chemical imbalances for a star with a certain value of 
$\Omega \dot{\Omega}$.
Figure \ref{fig:4} shows the solution to this equations when
only the neutrons are superfluid, such that $\Delta_{thr}=\Delta_n$.
This illustrates that the chemical imbalance always stabilizes at a value 
slightly higher than $\Delta_{thr}$,
\begin{eqnarray}
 \eta_{npe}^{\infty,qs}\approx\eta_{np\mu}^{\infty,qs}\gtrsim\Delta_{thr}= \mbox{min}\left(  \Delta_n+3\Delta_p,3\Delta_n+\Delta_p \right).
\end{eqnarray}

We can also observe from  Fig.~\ref{fig:4} that the
solution becomes closer to  $\Delta_{thr}$  as we increase its value.
For higher values of the gap, this is because the chemical imbalances at the quasi-steady state are higher. 
For a given spin-down rate, the net reaction rate is then fixed and the 
nonsuperfluid reactions become more efficient, increasing as $\sim\eta^7$. Hence,
the reduction factors need to block more reactions, which happens 
for values of the chemical imbalance closer to $\Delta_{thr}$.
We show in Fig.~\ref{fig:9} (Appendix \ref{sec:B}) that the reduction factors indeed
become smaller as the chemical imbalance approaches the energy gap
from above. 

\subsubsection{Analytical approximation}\label{sec:results_qs_aprox}

The net reaction rates $\Delta\tilde{\Gamma}_{npe}$ and $\Delta\tilde{\Gamma}_{np\mu}$
are determined entirely by Eqs. (\ref{eq:equilibrio_eta1}) and (\ref{eq:equilibrio_eta2}).
Solving these equations, we obtain
\begin{eqnarray}
\label{eq:gamma_e}
\Delta\tilde{\Gamma}_{npe}&=&2 \Omega \dot{\Omega}  \frac{Z_{np\mu} W_{npe}-Z_{np} W_{np\mu}}
{Z_{npe}Z_{np\mu}-Z_{np}^2}\equiv -K_{e} \Omega \dot{\Omega}, \\
\label{eq:gamma_mu}
\Delta\tilde{\Gamma}_{np\mu}&=&2 \Omega \dot{\Omega}  \frac{Z_{npe} W_{np\mu}-Z_{np} W_{npe}}
{Z_{npe}Z_{np\mu}-Z_{np}^2}\equiv -K_{\mu} \Omega \dot{\Omega},
\end{eqnarray}
where the constants $K_{e}$ and $K_{\mu}$ depend exclusively on the
stellar structure, i.e., the EOS and the stellar mass (see \citealt{FR05} for 
the definition of these constants).

In the quasi-steady state, in which $\xi_l\gtrsim10^2$, the polynomials in Eqs. (\ref{eq:F_M_def}) and 
(\ref{eq:H_M_def}) are in a very good approximation $F(\xi_l)=\frac{9\xi_l^8}{11513\pi^8}$ and $H(\xi_l)=\frac{24\xi_l^7}{11513\pi^8}$, respectively. 
Thus, from the definitions (\ref{eq:L_noneq}), (\ref{eq:gamma_noneq}), (\ref{eq:R_epsilon}), 
(\ref{eq:R_gamma}), and doing some algebra
 we substitute the expressions of Eqs. (\ref{eq:gamma_e}) and (\ref{eq:gamma_mu}) to Eq. (\ref{eq:equilibrio_Ti}) 
to obtain
 \begin{eqnarray}
\label{eq:L_approx1}
& &L_\gamma^{\infty,qs}  =\left[ \eta_{npe}^{\infty,qs} K_e\left(1-\frac{9}{24}
		   \frac{\tilde{L}_{ne} R_{M,\epsilon}^n\left(\eta_{npe}^{\infty,qs}\right)
			+\tilde{L}_{pe} R_{M,\epsilon}^p\left(\eta_{npe}^{\infty,qs}\right)}
			{\tilde{L}_{ne} R_{M,\Gamma}^n\left(\eta_{npe}^{\infty,qs}\right)+
			\tilde{L}_{pe} R_{M,\Gamma}^p\left(\eta_{npe}^{\infty,qs}\right)}\right)+
\right.\nonumber \\  
  & & 		\left. \eta_{np\mu}^{\infty,qs}K_\mu\left(1-\frac{9}{24}
		   \frac{\tilde{L}_{n\mu} R_{M,\epsilon}^n\left(\eta_{np\mu}^{\infty,qs}\right)
			+\tilde{L}_{p\mu} R_{M,\epsilon}^p\left(\eta_{np\mu}^{\infty,qs}\right)}
			{\tilde{L}_{n\mu} R_{M,\Gamma}^n\left(\eta_{np\mu}^{\infty,qs}\right)+
			\tilde{L}_{p\mu} R_{M,\Gamma}^p\left(\eta_{np\mu}^{\infty,qs}\right)}\right)\right]
\left|\Omega \dot{\Omega}\right|,
\end{eqnarray}
where we have just rearranged the quasi-steady 
Eqs. (\ref{eq:equilibrio_Ti}), (\ref{eq:equilibrio_eta1}), and 
(\ref{eq:equilibrio_eta2}), and no approximation has yet been made.

 \begin{figure}[!h]
 \centering
\includegraphics[width=9cm,height=6cm]{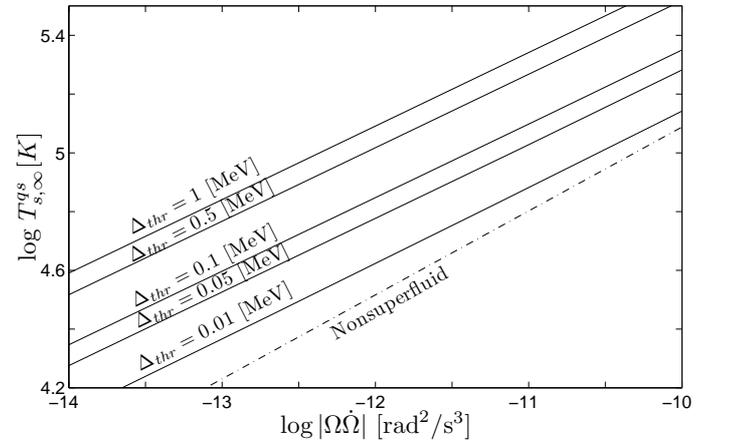}
\caption{ Surface temperature predicted for the quasi-steady state
for several values of 
$\Delta_{thr}=\min \left(\Delta_n+3\Delta_p,3\Delta_n+\Delta_p \right)$
(solid lines) and the nonsuperfluid case (dash-dotted line)
as a function of the spin-down rate $\Omega \dot{\Omega}$ with the same stellar
parameters as in Fig.~\ref{fig:4}. 
}
\label{fig:5}
\end{figure}

The first approximation that we make, as previously discussed, is to consider that 
the chemical imbalances are 
$\eta_{npe}^{\infty,qs}\approx\eta_{np\mu}^{\infty,qs}\approx\Delta_{thr}^\infty$.
This limit is valid when the energy gaps are relatively large, as 
we illustrated in Fig.~\ref{fig:4}. We checked this assumption for several solutions
of the rotochemical heating evolution equations and found that it is 
a reasonable approximation for $\Delta_{thr}>0.05$ MeV.

Our second approximation is to neglect the ratios of the 
reduction factors $R_{M,\epsilon}^N/R_{M,\Gamma}^N$. As we  
discussed in Sect. \ref{sec:theory_emiss_red}, the emissivity is more suppressed by the gaps than the 
net reaction rate when $\eta>\Delta$. It becomes an acceptable limit when the values of the 
gaps are relatively large compared to the thermal energy, say $\Delta/kT\gtrsim10^2$, which
is the limit of interest for rotochemical heating.
We checked that when we consider $\Delta_{thr}\sim0.05-1$ MeV, 
the ratio is $R_{M,\epsilon}^N/R_{M,\Gamma}^N\lesssim0.1$.

The first simplification tends to increase the predicted luminosity, while
the second approximation tends to decrease it. Thus, both effects together tend to
balance each other.
Finally, placing both approximations together, we obtain the simple expression 
for the bolometric luminosity of
 \begin{eqnarray}
\label{eq:L_approx2}
L_\gamma^{\infty,qs} & \approx&\left( K_e+K_\mu  \right)\Delta_{thr}\left|\Omega \dot{\Omega}\right|,
\end{eqnarray}
whose  physical interpretation is that the spin-down compression
sets the number of reactions per unit time and each one of these reactions releases an amount
of energy $\Delta_{thr}$ to reheat the star, as in the 
description of \citet{reis97}. 
The approximation is even more accurate for the surface temperature, since 
 $T_{s,\infty}^{qs}\propto \left(L_\gamma^{\infty,qs}\right)^{1/4}$
and the relative error in the lumimosity will lead to a smaller relative error
in the surface temperature.

In Fig.~\ref{fig:5}, we show the results of the predicted
surface temperature in the quasi-steady state as a function of the spin-down rate
by solving the equilibrium equations without approximations and 
using the black-body law to relate $L_\gamma^{\infty,qs}$ and $T_{s,\infty}^{qs}$. 
From this plot, we can infer that the curves are parallel and 
the surface temperature depends on some power of the spin-down that differs
from the power of the nonsuperfluid case, which is higher.
We can explain this behavior using the approximate expression 
$T_{s,\infty}^{qs}\propto |\Omega \dot{\Omega}|^{1/4}$,
while for the nonsuperfluid case,  \citet{FR05} obtain 
$T_{s,\infty}^{qs}\propto |\Omega \dot{\Omega}|^{2/7}$.

\begin{figure}
\centering
\includegraphics[width=9cm,height=6cm]{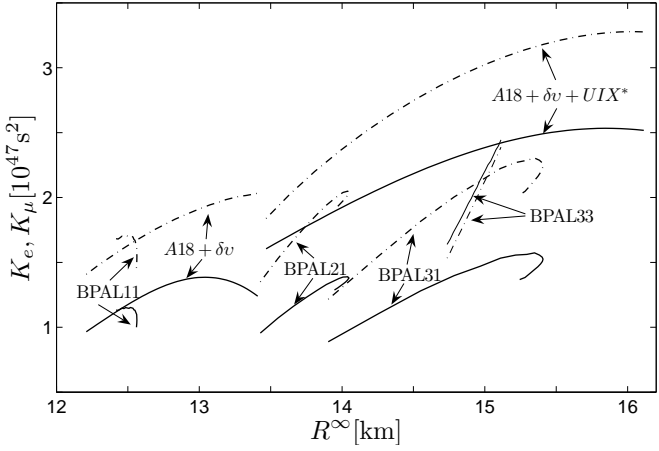} 
\caption{ Values of $K_e$ (solid lines) and $K_\mu$
(dash-dotted lines) for different
EOSs for a range of masses from 1 M$_\odot$ to the mass at which 
direct Urca with electrons is opened for each EOS.
}
\label{fig:6}
\end{figure}
 
In Fig.~\ref{fig:6}, we plot $K_e$ and $K_\mu$ as function of the radius for two 
sets of EOSs.
First, we show the two more realistic EOSs from \citet{apr98}
for the core (A18 + $\delta \upsilon$ and A18 + $\delta \upsilon$ + UIX*), 
supplemented with that of \citet{prl95} and \citet{hanpichon94} for the inner and outer
crust, respectively.
Second, we plot four representative EOSs from \citet{pal88}, which open direct Urca
reactions for stellar masses higher than 1 M$_\odot$. 
For this set of EOSs, we show in Fig.~\ref{fig:6} that $K_e+K_\mu\simeq(2-6)\times10^{47}\left[\mbox{s}^2\right]$, which, 
using Eq. (\ref{eq:L_approx2}), infers a bolometric luminosity of
 \begin{eqnarray}
\label{eq:L_approx3}
L_\gamma^{\infty,qs} & \simeq& \left(1-4\right) \times 10^{32}\left(\frac{\Delta_{thr}}{\mbox{MeV}}\right)
\left(\frac{\dot{P}_{-20}}{P_{\mbox{ms}}^3}\right)\mbox{ erg}\mbox{ s}^{-1},
\end{eqnarray}
where $\dot{P}_{-20}$ is the period derivative in units of $10^{-20}$ and
$P_{\mbox{ms}}$ is the period in milliseconds.
Finally, we can express the surface temperature using the expression for the luminosity 
given by Eq. (\ref{eq:L_approx2}) as
\begin{eqnarray}
\label{eq:T_approx}
T_{s,\infty}^{qs} & \simeq& \left(\frac{ K_e+K_\mu}{4\pi\sigma (R^{\infty})^2}\right)^{1/4}\Delta_{thr}^{1/4}\left|\Omega \dot{\Omega}\right|^{1/4},
\end{eqnarray} 
where $\sigma$ is the Stefan-Boltzmann constant. For the set of
EOSs that we use, we obtain 
 \begin{eqnarray}
\label{eq:T_approx2}
T_{s,\infty}^{qs} & \simeq& \left(5.7-6.6\right) \times 10^{5}
\left(\frac{\Delta_{thr}}{\mbox{MeV}}\right)^{1/4}
\left(\frac{\dot{P}_{-20}}{P_{\mbox{ms}}^3}\right)^{1/4}\mbox{ K}.
\end{eqnarray}

\subsubsection{Constraints on the energy gaps}\label{sec:results_qs_constraint}

\begin{figure}
\centering
\includegraphics[width=9cm]{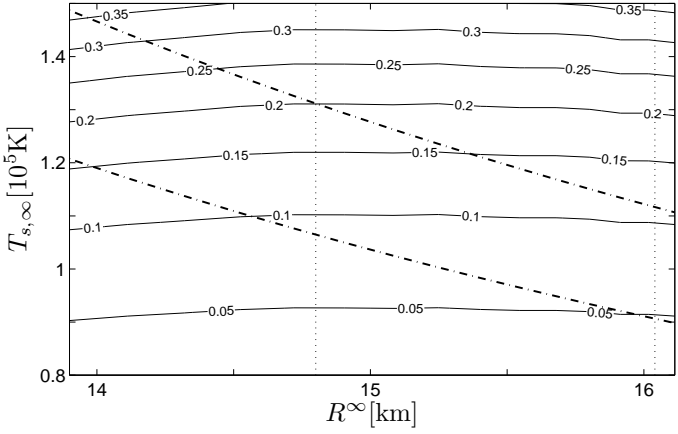} 
\caption{Surface temperature predicted in the quasi-steady state 
as a function of the the stellar radius for a star built with 
the A18 + $\delta \upsilon$ + UIX* EOS
for several values of $\Delta_{thr}$ in units of 
MeV (solid lines)  and the current spin-down rate of the MSP J0437.
Vertical dotted lines indicates the mass range $1.76\pm0.2$ M$_\odot$ 
allowed for MSP J0437 \citep{verbiest}.
Dashed-dotted lines are the  90$\%$ 
confidence contours of the black-body fit of
\citet{kargaltsev04} to probable thermal emission from this pulsar 
corrected for the latest distance of 157 pc \citep{deller}.
}
\label{fig:7}
\end{figure}

To explain the thermal emission of PSR J0437, as we showed in Fig.~\ref{fig:3}, it is necessary
 to invoke superfluidity, and the required values of the 
 energy gaps then fall in a  theoretically interesting range.
In Fig.~\ref{fig:7}, we compare the observational allowed range of surface 
temperatures \citep{kargaltsev04} with the theoretical predictions
for different values of $\Delta_{thr}$, using one particular
EOS. In this case, we obtain the allowed range
\begin{eqnarray}
\label{eq:constraint}
0.05 [\mbox{MeV}]\lesssim\mbox{min}\left(  \Delta_n+3\Delta_p,3\Delta_n+\Delta_p \right)\lesssim0.2 [\mbox{MeV}].
\end{eqnarray}
By adding two more EOSs (BPAL 21, BPAL 9) that forbid direct Urca reactions
in the allowed mass range, we expand the constraint to
\begin{eqnarray}
\label{eq:constraint_2}
0.05 [\mbox{MeV}]\lesssim\mbox{min}\left(  \Delta_n+3\Delta_p,3\Delta_n+\Delta_p \right)\lesssim0.45 [\mbox{MeV}].
\end{eqnarray}
These results depend of course on the assumption of 
spatially uniform and isotropic energy gaps. In principle, these are unrealistic
assumptions because the energy gaps depend on the local density and therefore
on the radius of the star, and the neutrons in the core are expected 
to form anisotropic Cooper pairs (see Sect. \ref{sec:theory_cooper}).

\subsubsection{Condition for arrival at the quasi-steady state}
\label{sec:results_qs_time}

To estimate the time taken for the star to arrive at the quasi-steady
state, we consider that the chemical imbalances grow
by the effect of $\Omega \dot{\Omega}$ in a unimpeded way, because the reactions
are highly suppressed, until they overcome the threshold imposed by the gaps.
Thus, the chemical imbalance associated with the lepton $l$ evolves as 
\begin{eqnarray}
\label{eq:eta_1_time}
\dot{\eta}^\infty_{npl} &=& 2W_{npl}\Omega \dot{\Omega},
\end{eqnarray}
until it exceeds the threshold imposed by gaps, at the moment 
the quasi-steady state is reached. Thus, we set the condition for 
arrival at the quasi-steady state as $\eta^\infty_{npl}=\Delta_{thr}$, 
which can be seen to be a good approximation from the lower panel
of Fig.~\ref{fig:3}.
Then, integrating the expression (\ref{eq:eta_1_time}) over time from the initial
spin period $P_i$ to the current value of the spin period $P$ and using the previous 
condition, we obtain the upper limit to the initial spin period of
\begin{eqnarray}
\label{eq:spin_condition2}
P_i < \left(\frac{1}{P^2} + \frac{\Delta_{thr}}{4 \pi^2\left|W_{npl}\right|}\right)^{-1/2} \approx
\left(\frac{1}{P_{ms}^2} + \frac{1}{4} \left(\frac{\Delta_{thr}}{\mbox{MeV}}\right)\right)^{-1/2} \mbox{ ms},
\end{eqnarray}
where $P_{ms}$ is the current value of the spin period in milliseconds.

For the case of PSR J0437, in particular,  we conclude that the initial
spin period to reach the quasi-steady state with the upper limit 
of the constraint in Eq. (\ref{eq:constraint_2}), i.e., $\Delta_{thr}=0.45$ MeV,
should be $P_i\lesssim2.7$ ms. 
This value could in principle be compared with the initial
period constraints from the cooling age of its white dwarf companion.
However,  this constraint is highly uncertain \citep{hansen98},
and we are therefore unable to draw definite conclusions.

\section{Conclusions}\label{sec:conclusion}

We have studied the rotochemical heating of millisecond pulsars 
with only modified Urca reactions in the presence of
uniform and isotropic Cooper pairing gaps for one or two nucleon species.

Based on these assumptions, we have found that the chemical imbalances in the star grow to the 
threshold value $\Delta_{thr}=\mbox{min}\left(  \Delta_n+3\Delta_p,3\Delta_n+\Delta_p \right)$,
which is higher than in the quasi-steady
state achieved in the absence of superfluidity. Therefore, the superfluid MSPs
will take longer to reach the quasi-steady state than their nonsuperfluid
counterparts, and they will have a higher luminosity in this state, given
by
 \begin{eqnarray}
\label{eq:L_approx4}
L_\gamma^{\infty,qs} & \simeq& \left(1-4\right) \times 10^{32}\left(\frac{\Delta_{thr}}{\mbox{MeV}}\right)
\left(\frac{\dot{P}_{-20}}{P_{\mbox{ms}}^3}\right)\mbox{ erg}\mbox{ s}^{-1}.
\end{eqnarray}
The constraint that we found for the energy gaps 
using our predicted effective temperatures and 
the black-body fit of \citet{kargaltsev04} to the UV emission of PSR J0437-4715 is
\begin{eqnarray}
0.05 [\mbox{MeV}]\lesssim\Delta_{thr}\lesssim0.45 [\mbox{MeV}].
\end{eqnarray}
In this sense, rotochemical heating presents an interesting 
tool for constraining the superfluidity parameters.
In a future paper, we will include the density dependence
of the energy gaps via the predictions of different
theoretical models.

\begin{acknowledgements}
We thank  Claudio Dib, Guillermo Cabrera, 
M\'arcio Catelan, Olivier Espinosa, Rafael Benguria, and 
Sebasti\'an Reyes,  for discussions and 
comments that benefited the present paper, an anonymous referee
for comments that improved the final version of this paper,
and Rodrigo Fern\'andez for letting us use his  
rotochemical heating code.
This work was supported by Proyecto Regular FONDECYT
1060644, the FONDAP Center of Astrophysics (15010003),
and Proyecto Basal PFB-06/2007.
\end{acknowledgements}

\onecolumn

\appendix


\section{Gauss-Laguerre integration method}\label{sec:A}
We describe the numerical method used in this work to calculate
the phase-space integrals involved in the net reaction rate and emissivity of 
the beta processes considering superfluid nucleon species.
\cite{villain} showed that Cooper pairing reduces these reactions and 
that there is no analytical expression to compute this suppression. 
These authors, therefore, solved  the phase-space integrals numerically using the so-called 
Gauss-Legendre quadrature, logarithmic variables, and  cut-off
values to cover a wide range of the unbounded integration domain.
In this work, it was more convenient for us to use the method presented here 
because it involves fewer evaluations
than the method used in \citet{villain} for a reasonable precision.

The method we show  is motivated by the behavior of the integrand 
for the beta processes, which decays exponentially
with each variable because of the Fermi functions. 
This happens asymptotically even when some particles are superfluid. For this reason,
among the Gaussian quadratures, a natural candidate for an interpolation
polynomial are the so-called Laguerre polynomials because these are defined on the 
basis of the continuous functions in $[0, \infty]$, whose inner product is defined with 
weight function $W(x)=e^{-x}$ \citep{abramo}. 
If the function to be integrated is the product of an $n$th-order polynomial
and an exponential function, this method is exact using an nth-order Laguerre polynomial
in the Gaussian quadrature. 
Therefore, a necessary condition for this method to work properly for an integral 
$\int_0^{\infty}f(x)dx$  is that $f(x)\cdot W(x)$ has to be a smooth
function, such as a polynomial \citep{davis}.

We present how this method has been applied to the phase-space integrals involved
in the net reaction rate for the direct Urca process to save notation. 
An extension to the Modified Urca processes is straightforward by adding two more superfluid
variables to the subsequent expressions. 
Considering the limits in the positive domain of each integration variable
the integral is given by
\begin{eqnarray}
\label{eq:durca_no_aprox_B}
 I_{D,\Gamma}&=& \int_0^{\infty} dx_{\nu} x_{\nu}^2 \int_{0}^{\infty}\int_{0}^{\infty}
	  dx_{n}dx_{p}\times \left\{ \sum_{j_n=\pm 1} \sum_{j_p=\pm 1}  f\left(j_nz_n\right) 
	 f\left(j_p z_p\right)              
\times       \left[ f\left(x_{\nu}+\xi-j_n z_n-j_p  z_p\right)
  	   -f\left(x_{\nu}-\xi-j_n z_n-j_p  z_p\right) \right] \right\},
\end{eqnarray}
where $z_i=\sqrt{x_i^2+\delta_i^2}$ ($i=n,p$) and 
$f(\cdot)$ is the Fermi function. 
In what follows, we argue that this integral satisfies the Laguerre quadrature
condition explained above, and we then express the numerical formula to 
compute it.

The integrand is an exponentially decaying function of the neutrino variable, 
and thus, it satisfies the quadrature condition. 
The integrand for the nucleons  has the shape of a multiplication of 
two Fermi functions. For instance, if we consider the neutron variable 
we have $f(\pm\sqrt{x_n^2+\delta_n^2})\cdot f(\alpha \mp \sqrt{x_n^2+\delta_n^2})$,
where $\alpha$ depends on the neutrino variable, the proton variable, and the chemical imbalance.
It is easy to see that the asymptotic behavior of this function is exponential by separating 
both cases and imposing the limiting case $x_n \gg \delta_n$, which is the relevant scale for this
variable. Thus, in this limit  $f(-\sqrt{x_n^2+\delta_n^2})\sim e^{-x_n}$ and 
$f(\sqrt{x_n^2+\delta_n^2})\sim 1$, while the other Fermi functions have the opposite
behaviors $f(\alpha - \sqrt{x_n^2+\delta_n^2})\sim 1$ and 
$f(\alpha + \sqrt{x_n^2+\delta_n^2})\sim e^{-x_n}$, respectively. 
Therefore, the nucleons also satisfy the quadrature condition, which may, however,
also be applicable to higher values
of these variables depending on the values that $\alpha$ take. In this sense,
higher values of $\alpha$, which depends on the inputs $\delta_n$, $\delta_n$,
and $\xi$, will make the quadrature less accurate for a given amount 
of point evaluations.

An additional step in applying this quadrature method is to define a 
relevant scale for each integration variable, as the scale of
 the argument of the Fermi function $f\left(x_{\nu}\pm \xi-j_n z_n-j_p z_p\right)$ 
 in the integrand of Eq. (\ref{eq:durca_no_aprox_B}), as in \citet{villain}.
For the neutrino variable, this scale is $S_{x_\nu}\equiv \xi+\delta_n+\delta_p$, and for the 
superfluid nucleons they are  $S_{x_n}\equiv \xi+\sqrt{2\delta_n}+\delta_p$ and
$S_{x_p}\equiv \xi+\sqrt{2\delta_p}+\delta_n$. 
The numerical formula is, finally
\begin{eqnarray}
\label{eq:lag_Durca}
 I_{D,\Gamma}&=&  \sum_{i_{\nu}=1}^{n_{\nu}} \sum_{i_n=1}^{n_n} \sum_{i_p=1}^{n_p} \sum_{j_n=\pm1} \sum_{j_p=\pm1}  S_{x_\nu}S_{x_n}S_{x_p} x_{i_\nu}^2\cdot 
 f\left(j_n z_{i_n}\right)  f\left(j_p z_{i_p}\right) 
\times \left[f\left(x_{i_\nu}+\xi-j_n z_n-j_p z_{i_p}\right)-
f\left(x_{i_\nu}-\xi-j_n z_{i_n}-j_p z_{i_p}\right)\right],
\end{eqnarray}
where $z_{i_{n}}=\sqrt{x_{i_{n}}^2+\delta_{n}^2}$ and 
$z_{i_{p}}=\sqrt{x_{i_{p}}^2+\delta_{p}^2}$. 
The values $x_{i_{\nu}}$, $x_{i_{n}}$, and $x_{i_{p}}$ are the roots of the Laguerre 
polynomials of order $n_\nu$, $n_{n}$, and $n_{p}$, 
multiplied by $S_{x_\nu}$, $S_{x_n}$, and $S_{x_p}$, respectively .
The values $W_{i_{\nu}}$, $W_{i_{n}}$, and $W_{i_{p}}$ are the weight factors
of the Laguerre polynomials, which can easily be  
obtained from tables or using their usual recursive formulae.

In practice, this method is used for Modified Urca processes where there are four
superfluid nucleon variables and the integration dimensions increase to five.

\begin{figure}[h!]
   \centering
\includegraphics[width=11cm]{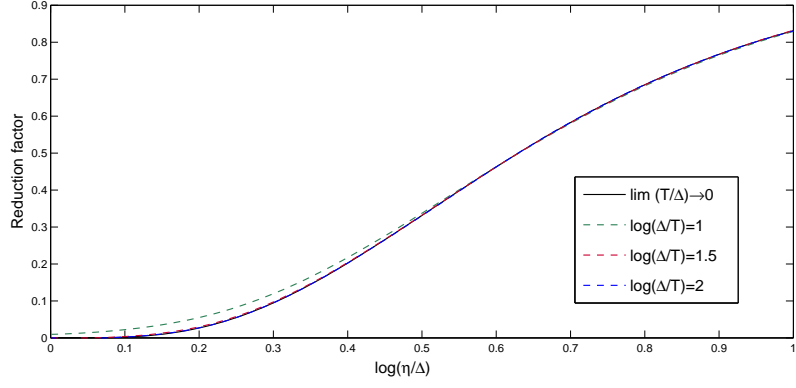}
\caption{ Relative error in the integral for the Legendre quadrature and Laguerre 
quadrature as a function of the number of evaluations considered in each
dimension for the integral of the direct Urca net reaction rate 
with $\xi=\delta_n=\delta_p=10$.
}
\label{fig:8}
\end{figure}

In Fig.~\ref{fig:8}, we provide an example where the Laguerre quadrature 
is more accurate than the Legendre quadrature over a few evaluations. However,
to achieve high precision the Legendre-like integration method is 
preferable because it can be used to perform many evaluations raising the number 
of point evaluations, while it is more difficult to compute the 
roots and weights of the Laguerre method using recursive methods to high orders of these 
polynomials.

\section{Analytical approximation}\label{sec:B}
In the context of rotochemical heating,  most of the MSP's lifetime 
is in a regime where the chemical imbalance is much higher 
than the temperature ($\xi=\frac{\eta}{kT}\gtrsim100$). 
On invoking superfluidity, the chemical imbalances increase
until they become of the order of the superfluidity energy gaps.
Thus, a relevant case to analyze and calculate in this work is the limit
\begin{equation}
 \xi \sim  \delta_{n,p}=\frac{\Delta_{n,p}}{kT}\gg1.
\end{equation}
The first analysis is given for the simplest case: the direct Urca process. 
The idea is to extend this analysis to the most general and complicated case:
 modified Urca with two superfluid nucleons.
After integrating over the electron variable, the phase-space integral for the net reaction rate is
\begin{eqnarray}
\label{eq:durca_no_aprox}
 I_{D,\Gamma}&=&\int_0^{\infty}dx_{\nu} x_{\nu}^2 \int_{-\infty}^{\infty}\int_{-\infty}^{\infty}
dx_{n}dx_{p}f(z_n)f(z_p)\times \{f(x_{\nu}+\xi-z_n-z_p) - f(x_{\nu}-\xi-z_n-z_p) \},
\end{eqnarray}
where $z_{i}=\mbox{sgn}(x_{i})\sqrt{x_{i}^2+\delta_{i}^2}$, with $i={n,p}$. 

In the regime of interest, the Fermi functions behave almost like
a step function, since the thermal scale is negligible compared
to the other relevant scales.
Moreover, one of the conclusions
of \citet{villain}, obtained by means of their numerical results,
is that for large enough energy gaps, say $\log(\Delta/T)\geq 1$, and as soon
as $\eta>\Delta$, the reduction factors of the net reaction rates for 
direct Urca  and modified Urca reactions  become very nearly independent of the temperature. 
This motivates us to take the limit of zero temperature.
Thus, replacing the Fermi functions by step functions, i.e.,
 $f(x)\longrightarrow \Theta(-x)$, the integral gives
\begin{eqnarray}
\label{eq:durca_no_aprox}
 I_{D,\Gamma}&=&\int_0^{\infty}dx_{\nu} x_{\nu}^2 \int_{-\infty}^{\infty}\int_{-\infty}^{\infty}
dx_{n}dx_{p}\Theta(-z_n)\Theta(-z_p)\times \{\Theta(-x_{\nu}-\xi+z_n+z_p) - \Theta(-x_{\nu}+\xi+z_n+z_p) \}.
\end{eqnarray}
Rearranging the limits of integration and eliminating the 
$\mbox{sgn}(\cdot)$ function from $z_{i}=\mbox{sgn}(x_{i})\sqrt{x_{i}^2+\delta_{i}^2}$,
 the integral can be reduced to
\begin{eqnarray}
\label{eq:durca_aprox}
 I_{D,\Gamma}&=&\int_0^{\infty}\int_0^{\infty}\int_0^{\infty}dx_{\nu}
dx_{n}dx_{p} x_{\nu}^2 \Theta\left(\xi-x_{\nu}-\sqrt{x_n^2+\delta_n^2}-\sqrt{x_p^2+\delta_p^2}\right).
\end{eqnarray}
By applying these approximations, the reactions can be seen to be open only when
$\delta_{n}+\delta_{p}<\xi$. (see  \citealt{reis97} for a schematic justification).

The integral in Eq. (\ref{eq:durca_aprox}) appears not to have a general, closed expression in terms of the three 
parameters $\xi, \delta_n, \delta_p$, but it is possible to perform two out of 
three variable integrations. By considering only one superfluid 
nucleon, an analytical expression can be found. In what follows, 
we explain these results.

The first step is to define an appropriate change of variables
\begin{eqnarray}
\label{eq:cambio_xi}
 u_{\nu}\equiv \frac{x_{\nu}}{\xi} \mbox{ } \mbox{  and  } \mbox{ }  u_{i}\equiv 
\sqrt{ \frac{x_{i} ^2+\delta_{i}^2}{\xi^2}} \mbox{,   } (i=n,p) \mbox{,}
\end{eqnarray}
where the Jacobian of this transformation is $J(u_{\nu},u_n,u_p)=\xi^3\frac{u_nu_p}{\sqrt{u_n^2-(\Delta_n/\eta)^2}\sqrt{u_p^2-(\Delta_p/\eta)^2} }$, 
considering that $\Delta_p/\eta=\delta_p/\xi$, and the new limits of integration 
must satisfy the upper limit imposed by the step function
and the lower limit defined by the change of variables. 
Thus, the new integral is 
 \begin{eqnarray}
\label{eq:aprox_final_neutrino}
I_{D,\Gamma} &=&\xi^5\int_{\Delta_n/\eta}^{1}\int_{\Delta_p/\eta}^{1-u_n}\int_0^{1-u_n-u_p} du_{n} du_{p} du_{\nu} u_{\nu}^2\frac{u_n u_p}{\sqrt{u_n^2-(\Delta_n/\eta)^2}\sqrt{u_p^2-(\Delta_p/\eta)^2} }.
\end{eqnarray}
Integrating over the neutrino variable, we obtain
 \begin{eqnarray}
\label{eq:aprox_final_neutrino2}
I_{D,\Gamma} =\frac{\xi^5}{3}\int_{\Delta_n/\eta}^{1}\int_{\Delta_p/\eta}^{1-u_n} du_{n} du_{p} (1-u_n-u_p)^3\frac{u_n u_p}{\sqrt{u_n^2-(\Delta_n/\eta)^2}\sqrt{u_p^2-(\Delta_p/\eta)^2} }.
\end{eqnarray}
The integration over the proton variable yields
 \begin{eqnarray}
\label{eq:aprox_final_proton}
I_{D,\Gamma} &=& \frac{\xi^5}{3}\int_{\Delta_n/\eta}^{1-\Delta_p/\eta} du_{n}\frac{u_n}{\sqrt{u_n^2-(\Delta_n/\eta)^2}} K_{\Gamma}(u_n,\Delta_p/\eta) \mbox{  }\mbox{ ,} 
\end{eqnarray}
where
 \begin{eqnarray}
\label{eq:K_gama}
K_{\Gamma}(u_n,\Delta_p/\eta)&=&-\frac{3(\Delta_p/\eta)^2}{2}\left[\frac{(\Delta_p/\eta)^2}{4} +(1-u_n)^2 \right]
\cdot \ln\left[\frac{1-u_n+\sqrt{(1-u_n)^2-(\Delta_p/\eta)^2}}{\Delta_p/\eta}\right] \\ \nonumber
&+&\sqrt{(1-u_n)^2-(\Delta_p/\eta)^2}
\left[\frac{13(\Delta_p/\eta)^2}{8}\left(1-u_n \right) + \frac{\left(1-u_n \right)^3}{4}  \right].
\end{eqnarray}

The latter integral in Eq. (\ref{eq:K_gama}) appears not to have an analytical 
closed formula. For this reason, the solution hereafter is taken numerically. 
The method for the emissivity integral is completely 
analogous, but with the only difference that we replace $x_\nu^2$ with
$x_\nu^3$ in Eq. (\ref{eq:durca_aprox}).
The formula obtained in this case is
 \begin{eqnarray}
\label{eq:aprox_final_proton_emis}
I_{D,\epsilon} &=& \frac{\xi^6}{4}\int_{\Delta_n/\eta}^{1-\Delta_p/\eta} du_{n}\frac{u_n}
{\sqrt{u_n^2-(\Delta_n/\eta)^2}} K_{\epsilon}(u_n,\Delta_p/\eta) \mbox{  }\mbox{ ,} 
\end{eqnarray}
where
\begin{eqnarray}
\label{eq:K_epsilon}
 K_{\epsilon}(u_n,\Delta_p/\eta)&=&2(\Delta_p/\eta)^2\left[\frac{3}{4}(\Delta_p/\eta)^2(1-u_n)+(1-u_n)^3\right]
\cdot \ln \left[\frac{ 1-u_n+\sqrt{(1-u_n)^2-(\Delta_p/\eta)^2}} {\Delta_p/\eta} \right]   \\ \nonumber
&+&  \frac{\sqrt{(1-u_n)^2-(\Delta_p/\eta)^2}}{5}
\left\{ (1-u_n)^4+\frac{(\Delta_p/\eta)^2}{6} \left[ 83(1-u_n)^2+16 (\Delta_p/\eta)^2\right] \right\}.
\end{eqnarray}

The final step is to extend the same reasoning to the modified Urca  
processes, and is natural on the basis of the previous analysis. 
Here  one has to distinguish between the \textit{neutron branch} (superscript $n$)  and 
the \textit{proton branch} (superscript $p$), but for the similarity relation 
 $I_{M,\Gamma}^{p}(\delta_n,\delta_p)=I_{M,\Gamma}^{n}(\delta_p,\delta_n)$ only
one expression is needed. 
Thus, without loss of generality, we only write the neutron branch expression of 
the net reaction rate of Murca
 \begin{eqnarray}
\label{eq:aprox_final_pro_MU}
I_{M,\Gamma}^{n}&=&\frac{\xi^7}{3}\int_{\Delta_n/\eta}^{1}\int_{\Delta_n/\eta}^{1-u_n}
\int_{\Delta_n/\eta}^{1-u_n-u_{n_i}}du_{n} du_{n_i} du_{n_f}
 \frac{u_n u_{n_i} u_{n_f}}
{\sqrt{u_n^2-(\Delta_n/\eta)^2}\sqrt{u_{n_i}^2-(\Delta_n/\eta)^2}\sqrt{u_{n_f}^2-(\Delta_n/\eta)^2}}  K_{\Gamma}(u_n+u_{n_i}+u_{n_f},\Delta_p/\eta),
\end{eqnarray}
where  $K_{\Gamma}(\cdot,\cdot)$ is the function given by Eq. (\ref{eq:K_gama})
and $n_i$ and $n_f$ represent the initial and final neutron spectator, respectively.
In the same way, the emissivity has an expression in terms of the function $K_{\epsilon}(\cdot,\cdot)$
defined by Eq. (\ref{eq:K_epsilon}):
 \begin{eqnarray}
\label{eq:aprox_final_pro_MU_emis}
I_{M,\epsilon}^n&=&\frac{\xi^8}{4}\int_{\Delta_n/\eta}^{1}\int_{\Delta_n/\eta}^{1-u_n}\int_{\Delta_n/\eta}^{1-u_n-u_{n_i}}du_{n} du_{n_i} du_{n_f}
  \frac{u_n u_{n_i} u_{n_f}}
{\sqrt{u_n^2-(\Delta_n/\eta)^2}\sqrt{u_{n_i}^2-(\Delta_n/\eta)^2}\sqrt{u_{n_f}^2-(\Delta_n/\eta)^2}} K_{\epsilon}(u_n+u_{n_i}+u_{n_f},\Delta_p/\eta).
\end{eqnarray}
In both cases, the net reaction rate and the emissivity, the condition to recover 
the non-superfluid case asymptotically is satisfied.

The advantage of this treatment is that the numerical integration of the net reaction 
rate and the emissivity for the Urca processes is much faster than those without 
these approximations. 
First, the region of integration is bounded, while those of the initial integrals 
are not. This reduces considerably the number of points used in each integral.
Secondly, the dimensions to integrate are reduced to one in the case of direct 
Urca and to three in the case of modified Urca.
In practice, to compute these integrals we use the \textit{tanh-sinh} method 
\citep{tanh}, which integrates the singularities in $\Delta_n/\eta$ and/or in 
$\Delta_p/\eta$ quite rapidly.

The integrals in Eqs. (\ref{eq:aprox_final_proton}), (\ref{eq:aprox_final_proton_emis}), 
(\ref{eq:aprox_final_pro_MU}), and (\ref{eq:aprox_final_pro_MU_emis}) 
can be solved analytically when only one superfluid reactant is present, which is
characterized by the energy gap $\Delta$. 
The reduction factor $R$ (for all the previous cases) obtained from this simplification 
is given by the integral
 \begin{eqnarray}
\label{eq:integral_una dim}
R &=& (\alpha+1) \int_{\Delta/\eta}^{1} du\frac{u (1-u)^\alpha}{\sqrt{u^2-(\Delta/\eta)^2}} \mbox{  }\mbox{ ,} 
\end{eqnarray}
where $\alpha=4$ and  $\alpha=5$ for Durca processes 
(net-reaction rate and emissivity, respectively), and
$\alpha=6$ and  $\alpha=7$ for Murca processes 
(net-reaction rate and emissivity, respectively).
The solution to this integral is
 \begin{eqnarray}
\label{eq:R_anal}
R(\Delta_n/\eta)&=&P(\Delta/\eta)\ln\left(\frac{1}{1+\sqrt{1-(\Delta/\eta)^2}}\right)
+Q(\Delta/\eta)\sqrt{1-(\Delta/\eta)^2},
\end{eqnarray}
where $P(\cdot)$ and $Q(\cdot)$ are the polynomials
 \begin{eqnarray}
\label{eq:poli_1}
P(x)&=&10x^2+ \frac{15}{2}x^4 \mbox{ } \mbox{ } \mbox{ and }\mbox{ } \mbox{ }
Q(x)=1+\frac{83}{6}x^2 +\frac{8}{3}x^4 \quad \mbox{for the Durca net reaction rate,}\\
\label{eq:poli_2}
P(x)&=&15x^2+ \frac{45}{2}x^4+ \frac{15}{8}x^6\mbox{ } \mbox{ } \mbox{ and }\mbox{ } \mbox{ }
Q(x)=1+\frac{97}{4}x^2 +\frac{113}{8}x^4 
\quad \mbox{for the Durca emissivity,}\\
\label{eq:poli_3}
P(x)&=&21x^2+ \frac{105}{2}x^4 + \frac{105}{8}x^6\mbox{ } \mbox{ } \mbox{ and }\mbox{ } \mbox{ }
Q(x)=1+\frac{759}{20}x^2 +\frac{1779}{40}x^4 +\frac{16}{5}x^6
\quad \mbox{for the Murca net reaction rate,}\\
\label{eq:poli_4}
P(x)&=&28 x^2+ 105 x^4 + \frac{105}{2}x^6 + \frac{35}{16}x^8\mbox{ } \mbox{ } \mbox{ and }\mbox{ } \mbox{ }
Q(x)=1+\frac{551}{10}x^2 +\frac{4327}{40}x^4+ \frac{1873}{80}x^6
\quad \mbox{for the Murca emissivity.}
\end{eqnarray}

\begin{figure}[h!]
   \centering
\includegraphics[width=10cm, height=5cm]{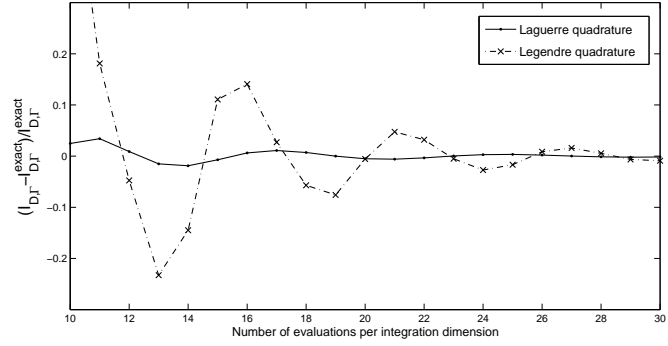}
\caption{Reduction factor $R$ for the net reaction rate of direct Urca with one superfluid 
particle as a function of the ratio $\eta/\Delta$ for several values of the
ratio $\Delta/T$ (dashed lines)  using the method detailed in Appendix \ref{sec:A}, and the 
limit case $T/\Delta\rightarrow0$ calculated from Eqs. (\ref{eq:R_anal}) and 
(\ref{eq:poli_1}) (solid line).
} 
\label{fig:9}
\end{figure}

Finally, we show  in Fig.~\ref{fig:9} a comparison between our zero temperature approximation and the exact 
calculations for finite values of temperature, which demonstrates the accuracy of our approximation
even for values of the energy gap compared that are not very large relative
to the temperature, say $\Delta/T\gtrsim30$.

\end{document}